\documentclass[12pt]{article}
\usepackage[pctex32]{graphics}
\begin{document}
\vspace{1cm}
\begin{center}
~
\\
~
{\bf  \Large Thermodynamics of  Ideal Boson and Fermion Gases in the Static Taub Universe}
\vspace{1cm}

                      Wung-Hong Huang\\
                       Department of Physics\\
                       National Cheng Kung University\\
                       Tainan, Taiwan\\

\end{center}
\vspace{1cm}
Some thermodynamic quantities of nonrelativistic ideal boson and fermion gases in the static Taub universe are derived to first order in a small anisotropy parameter $d$ which measuring the deformation from the spherical Einstein universe. They are used to investigate the problem of how the curvature anisotropy affects the thermodynamic behaviors of an ideal gas. It is found that, when the universe is in the oblate configuration (i.e., $d > 0$), the effect of curvature anisotropy is to increase the number of the fraction in the Bose-Einstein condensation and to decrease the fermion distribution function at low temperature. When the universe is in the prolate configuration (i.e., $d < 0$), the effects of curvature anisotropy on the thermodynamic quantities is contrary to that in the oblate configuration. The density matrix of a two particle system is evaluated and it is used to define the ``statistical interparticle potential" as an attempt to give a ``statistical interpretation" about the found thermodynamic behaviors. It is found that when the universe is in the oblate (prolate) configuration the curvature anisotropy will enhance (reduce) both the ``statistic 
attraction" among the bosons and ``statistical repulsion" among the fermions. It is 
expected that such a behavior will also be shown in the relativistic system. 
\vspace{2cm}
\begin{flushleft}
J. Math. Phys. 35 (1994) 3594
\\
*E-mail:  whhwung@mail.ncku.edu.tw\\
\end{flushleft}

%%%%%%%%%%%%%%%%%%%%%%%

\newpage
\section{Introduction}
The investigation of the thermodynamics of particle field in the curvature space-time is useful to understand the dynamics of the universe. Many efforts have been made to study the problem of Bose-Einstein condensation[1-3] in the static Einstein universe, [4-7] both in the nonrelativistic []4] and ultra-relativistic systems [5-7]. The condensation problem on other space-time such as the hyperbolic manifolds [8] or the Kaluza Klein universe [9] have also been studied. 

In the present article we will study some thermodynamic quantities of nonrelativistic ideal boson and fermion gases in the static Taub universe with a small anisotropy. Our main purpose is to see how the curvature anisotropy will affect the statistical properties of ideal boson and fermion gases. One basic reason for dealing with the static Tau space is that the complete eigenvalue 
spectrum of the Laplacian is known. This feature makes it mathematically tractable. There are several investigations of the physics in the static Taub universe [10],which include the quantum field effects [11]and finite temperature field theory [12]. We will present the analytic forms of some interesting thermodynamic quantities expanded to the first order in a small anisotropy parameter $d$ measuring the deformation from the spherical Einstein universe and thus several interesting physical properties can be read explicitly.

The mode space-time and associated eigenfrequencies of a massless scalar particle [10] are briefly summarized in Sec. II. In Sec. III we study the phenomena of Bose-Einstein condensation along the algorithm of Al'taie [4] and find that, in the oblate configuration (i.e., $d > 0$) the effect of curvature anisotropy is to increase the fraction of condensation. In Sec. IV we show that the effect of curvature anisotropy (if $d > 0$) is to decrease the fermion distribution function at low temperature. The expressions derived in Sets. III and IV also show that effects of curvature anisotropy on the thermodynamic quantities of an ideal gas for the universe in the prolate configuration (i.e., 
$d < 0$) are contrary to that in the oblate configuration. As an attempt to give a ``statistic interpretation" of those found thermodynamic behaviors we present in Sec. V an evaluation of the density matrix of a two-particle system and from this we define the ``statistical interparticle potential" along the idea of Uhlenbeck [13]. We find that in the oblate (prolate) configuration the effect of curvature anisotropy is to enhance (reduce) both the ``statistic attraction" among the bosons and ``statistical repulsion" among fermions. Section VI is devoted to a discussion while some mathematic calculations are collected in the two Appendices. 
For simplicity we only study in this paper the case of a spinless particle, however, the case of spin-l gas can be studied in a similar fashion. Note that although our investigation is confined to the nonrelativistic system, we expect that our main results of showing how the curvature anisotropy effects the thermodynamic behaviors will also be found in the relativistic system [14-16]. For example, the Bose-Einstein condensation is a low-temperature phenomena and the system at low temperature can be described by the nonrelativistic effects [14]. However, to confirm this one can extend the analysis into the relativistic regime. This is more difficult in mathematics. We hope to be able to report on this elsewhere. 

\section{Model} 
The background geometry of the Taub space has the metric 
$$ds2= -dt^2+ \sum_i\ell_i^2(\sigma^i)^2,\eqno{(2.1)}$$ 
where $\ell_1=\ell_2\ne\ell_3$ and the $\sigma^i$ form a basis one-form on $S^3$ satisfying the structure relations $d\sigma^i = {1\over2}\sigma^j\wedge\sigma^k$. In the Euler-angle parametrization the $\sigma^i$ are expressed as 
$$\sigma^1= cos\psi d\theta+ sin\psi sin \theta d\phi.$$
$$\sigma^2= -sin\psi d\theta+ cos\psi sin \theta d\phi.$$
$$\sigma^3= dpsi + cos \theta d\phi,\eqno{(2.2)}$$ 
where $0\le \theta\le\pi$,$0\le \psi, \phi\le 2\pi$. The $\ell_i's$ are the three principle curvature radii of the homogeneous space and are constants for a static universe. The case when all three $\ell_i's$ are equal is the closed 
Friedmann-Robertson-Walker universe. The volume and curvature scalar of the Taub space is given by 
$$V ={2\pi^2 a^3\over\sqrt{1+d}},\eqno{(2.3)}$$ 
$$R = {4\ell_1^2-\ell_3^2\over 4\ell_1^4} = {6(1+(4/3)d\over a^2(1+d)},\eqno{(2.4)}$$
where $a=2\ell$, and $d=(\ell_1^2/\ell_3^2)-1$ is an anisotropy parameter with range  $-1<d<\infty$. We call the configuration with $d >0$ oblate and that with $d < 0$ prolate.

 The eigenfrequencies of a massive (m) scalar field (4) with curvature coupling ($\xi$) to a static Taub universe are found to be [10]
$$\omega_L = \left[{J(J+1)\over \ell_1^2}+\left({1\over\ell_3^2}-{1\over\ell_1^2}\right)K^2 + m^2+{1\over 6}(1-\xi) R\right]^{1/2}. \eqno{(2.5)}$$
Here $L = {J,K,M}$: $J$ takes all values of positive integers and half integers and $K,M = -J, -J +1 , . . . , J - 1, J$. The quantum number $M$ is totally degenerate. The nonrelativistic approximation of Eq. (2.5) can be expressed as a n-degeneracy frequency 
$$\omega_{nq} = {1\over m a^2} [n^2+d(n-1-2q))^2],\eqno{(2.6)}$$
where $n = 1,2 ,...$, and $q =0,l,..., (n - 1)$. To obtain the above result we have discarded some overall constants which are irrelevant to our results. 
Notice that the eigenfrequency relation in the above equation, which is a nonrelativistic approximation and does not depend on the curvature coupling 6, can also be obtained by solving the nonrelativistic Schrodinger equation when the Laplacian operator is that on $S^3$ space defined by Eq. (2.2). Thus, we can regard Eq. (2.6) as the nonrelativistic approximation of the energy spectrum of an ideal gas, classical or quantum (boson or fermion), moving in the static Taub space and use it to investigate their thermodynamic behavior. 

\section{ Boson-Einstein Condensation} 
The behavior of a nonrelativistic ideal boson gas confined to background geometry of an Einstein universe has been investigated by Al'taie [4].  Apart from the expected effects of smoothing out the singularities of the thermodynamic functions at the transition temperature he also verified 
a displacement of the specific heat maximum towards higher temperature and an enhancement of the number of condensed particles due to the finite size of the system. This section is presented to prove that such an effect of a finite size will be enhanced (reduced) by the curvature anisotropy for the Taub universe in the oblate (prolate) configuration. 

The number of particles for an ideal Boson gas in our specific case is 
$$N = \sum_{n=1}^\infty \sum_{q=0}^{n-1}[exp[\beta(\omega_{nq}-\mu]-1]^{-1}\hspace{10cm}$$
$$=\sum_{n=1}^\infty \sum_{q=0}^{n-1}n[exp\{\beta'[n^2+d(n-1-2q)^2]+\alpha\}-1]^{-1}\hspace{6cm}$$
$$ =\sum_{n=1}^\infty\sum_{q=0}^{n-1}n[exp(\beta'n^2+\alpha)-1]^{-1} = d n(n-1-2q)^2 exp(\beta'n^2+\alpha)[exp(\beta'n^2+\alpha)-1]^{-2}$$
$$= \left[1+{d\over 3} \beta'\left({\partial\over\partial\beta'}-{\partial\over\partial\alpha}\right)\right]\sum_{n=1}n^2[expt(\beta'n^2)-1]^{-1},\hspace{5.6cm}\eqno{(3.1)}$$ 
where $\beta =1/kT$ and $\mu = -\beta^{-1} \alpha$ is the chemical potential. We have expanded the result to the first order in the anisotropy parameter $d$ and used the variables defined by 
$$\beta'={1\over 2mkTa^2 }= {1\over 4\pi} \left({\lambda\over a }\right)^2,~~~~\lambda = \left({2\pi\over mkT}\right)^{1/2},\eqno{(3.2)}$$
where $\lambda$ is the mean thermal wavelength of the particle. 

Substituting the Al'taie;s result [4] into the last summation in Eq. (3.1) we have
$$N = \left[1+{d\over 3} \beta'\left({\partial\over\partial\beta'}-{\partial\over\partial\alpha}\right)\right]\left[{1\over2}\left({1\over\beta'}\right)^{1/2}\left[\Gamma\left({3\over2}\right)F_{3/2}(\alpha)-2\pi\alpha^{1/2}S(2y) \right] \right], \eqno{(3.3)}$$
where 
$$F_\ell(\alpha) = \sum_{i=1}^\infty n^{-\ell}~e^{-n\alpha},~~~ S(2y) = {1\over e^{2y} -1},~~~y \equiv 2\pi^{3/2}\alpha^{1/2}\left({a\over\lambda}\right).\eqno{(3.4)}$$
The function $F_\ell(\alpha)$ has the properties $(\partial/\partial\alpha)F_\ell(\alpha) = - F_{\ell-1}(\alpha)$ and
$$F_{3/2}(\alpha) = \zeta \left({3\over2}\right) - 2 (\pi\alpha)^{1/2} + O(\alpha),~~~F_{1/2}(\alpha) = \pi^{1/2}\alpha^{-1/2}- \zeta \left({1\over2}\right) + O(\alpha).\eqno{(3.5)}$$
At low temperatures $\alpha << 1$, then using the above relations and after the calculation we obtain the result
$${N\over V}=\left({1\over \tilde\ell}\right)^3 \left\{\zeta\left({3\over2}\right)-{y\over \pi}\left({\lambda\over a}\right)cothy -d {\pi\over 6}\left({\lambda\over a}\right)\left[ {1\over y} cothy -{4\over\pi^2}(y^2+\pi^2) {e^{2y}\over(e^{2y}-1)^2}-{d\over12\pi}\left({\lambda\over a}\right)^2\zeta \left({1\over2}\right)\right]\right\},\eqno{(3.6)}$$
where $\tilde\ell$ is the mean interparticle distance. The term to the order of $(\lambda/a)^2$ will be neglected, hereafter, because $a>>\lambda$ in our case.
 
In the limit $a\rightarrow \infty$ and $d=0$, Eq. (3.6) becomes
$$\left({N\over V}\right)_{a\rightarrow \infty,~d=0}=\left({1\over \tilde\ell}\right)^3  =\left({1\over \lambda_0}\right)^3\zeta\left({3\over2}\right),\eqno{(3.7)}$$
This is the well-known bulk result. Using the definition of the mean thermal wavelength of the particle in Eq. (3.2) we have the relation  $x =T/T_0= (\lambda_0/\lambda)^2$, then from the Eqs. (3.6) and (3.7) we get the final result
$$\left({1 \over x}\right)^{3/2} = 1-{1\over\pi}\left[\zeta\left({3 \over 2}\right)  \right]^{-2/3}{\tilde \ell\over a}\left({1 \over x}\right)^{1/2}\left[ycothy-{d\over 6}\left[{\pi^2\over y}cothy-4(y^2+\pi^2){e^{2y}\over (r^{2y}-1)^2}\right]\right].{(3.8)}$$
The case of $d=0$ is that in Ref. 4. As the anisotropy parameter $d$ is small the ``function " $y$ can be expanded as 
$$ y=y_0+dy_1+O(d^2).\eqno{(3.9)}$$
and Eq. (3.8) can be reduced into two equations
$$ y_0cothy_0= {1\over\pi}\left[\zeta\left({3 \over 2}\right)  \right]^{-2/3}{\tilde \ell\over a}\left({1 \over x}\right)^{1/2}(1-x^{-3/2}).\eqno{(3.10)}$$
$$ y_1= {1\over 6(cothy_0-y_0csch^2y_0)}\left[{\pi^2\over y}cothy-4(y^2+\pi^2){e^{2y}\over (r^{2y}-1)^2}\right].\eqno{(3.11)}$$
 For a given $\tilde\ell/a$  and a given $x$ Eq. (3.10) has been used to find $y_0$ by Al'taie [4]. Once the $y_0$ is found the value of $y_1$  can easily be obtained from Eq. (3.11). Finally the value of $y$ is then used to evaluate the thermodynamic quantities. 

An interesting physical quantity is the condensate fraction $N_0/N$ [1-3]. The number of particles in the ground in our case is given by 
$$N_0 =[exp(\beta'+\alpha)-1]^{-1}=\left[exp\left({\pi^2\over y^2}+1\right)\alpha-1\right]^{-1}\approx {y^2\over \pi^2+y^2}{1\over\alpha}={y^2\over \pi^2+y^2}{4\pi^3\over y^2}\left({a\over\lambda)}\right)^3$$
$$ ={4\pi^3 x\over \pi^2+y^2}\left[\zeta\left({3\over2}\right)\right]^{3/2}\left({a\over \tilde\ell}\right) .\eqno{(3.12)}$$
Thus 
$${N_0\over N} = \left(1+{1\over2} d\right){2\pi x\over\pi^2+y^2}\left[\zeta\left({3\over2}\right)\right]^{3/2}\left({a\over \tilde\ell}\right) .\eqno{(3.13)}$$
Using the relation (3.9) the above equation is reduced to 
$${N_0\over N} = {N_0\over N} |_{a=0}+ \left({1\over\pi^2+y^2}\right)^2\pi x \left[\zeta\left({3\over2}\right)\right]^{3/2}\left({a\over \tilde\ell}\right) d [\pi^2+y_0^2-4y_0y_1].\eqno{(3.14)}$$

To see how the curvature anisotropy will effect the Bose-Einstein condensation one must know the behavior of the bracket term in the above equation. Although we have as yet no ability to prove it, the numerical evaluation shows that the bracket term is positive definite. Thus we conclude that when the universe is in the oblate configuration (i.e., $d>0$) the effect of curvature 
anisotropy is to increase the number of the fraction in the Bose-Einstein condensation. On the contrary, when the universe is in the prolate configuration (i.e., $d<0$) the effect of curvature anisotropy is to decrease the number of the fraction in the Bose-Einstein condensation. This behavior may be interpreted as that the increase (decrease) in the ``statistic attraction potential" among the boson gas by these effects, as shown in Sec. V, will push more (less) particles into the ground state. 
\section{Fermion Distribution Function}
The total number of the fermion particles in our model is derived from 
$$N = \sum_{\epsilon}[exp[\beta(\epsilon-\mu)]+1]\hspace{13cm}$$
$$ =\sum_{n=1}^\infty \sum_{q=0}^{n-1}[exp[\beta(\omega_{nq}-\mu]+1]^{-1}\hspace{10cm}$$
$$=\sum_{n=1}^\infty \sum_{q=0}^{n-1}n[exp\{\beta'[n^2+d(n-1-2q)^2]-\xi\}+1]^{-1}\hspace{6cm}$$
$$ =\sum_{n=1}^\infty\sum_{q=0}^{n-1}n[exp(\beta'n^2-\xi)+1]^{-1} = d n(n-1-2q)^2 exp(\beta'n^2+\alpha)[exp(\beta'n^2-\xi)+1]^{-2}$$
$$= \left[1+{d\over 3} \beta'\left({\partial\over\partial\beta'}+{\partial\over\partial\xi}\right)\right]\sum_{n=1}n^2[exp(\beta'n^2-\xi)-1]^{-1},\hspace{5.6cm}\eqno{(4.1)}$$ 
where $\mu=\beta^{-1}\xi$ is the chemical potential. We have expanded the result to the first order in the anisotropy parameter $d$ and used the variables defined in Eq. (3.2). 

The last summation in Eq. (4.1) is executed in Appendix A and we have 
$$N = \left[1+{d\over 3} \beta'\left({\partial\over\partial\beta'}+{\partial\over\partial\xi}\right)\right]\left[{1\over3}\left({\xi\over\beta'}\right)^{3/2}\left[1+{\pi^2\over8}{1\over\xi^2}+{3\over2}\left({\beta'\over\xi}\right)^{1/2} \right] \right]$$
$$ = {1\over3}\left({\xi\over\beta'}\right)^{3/2}\left(1-{d\over2}\right)\left[1+{\pi^2\over8}{1\over\xi^2}+{3\over2}\left(1+{d\over6}\right)\left({\beta'\over\xi}\right)^{1/2} \right]. \eqno{(4.2)}$$
The above equation leads to the result 
$${N\over V}= {g\over 6\pi^2}(2mkT\xi)^{3/2}\left[1+{\pi^2\over8}{1\over\xi^2}+{3\over4}\left({1\over\pi\xi}\right)^{1/2}\left({\lambda\over a}\right) \left(1+{d\over6}\right)\right],\eqno{(4.3)}$$
where g is a weight factor that arises from the internal structure of the fermion particle (such as the spin). The zeroth approximation of the above equation gives 
$$kT\xi_0 = {1\over 2m}\left({6\pi^2 N\over gV}\right)^{2/3}\epsilon_F,\eqno{(4.4)}$$
which is the ground-state result. Note that $\epsilon_F$ is a function of particle density $n=N/V$ which does not depend on the finite site radius, $a$, or on the curvature anisotropy, $d$. In the next approximation we obtain 
$$\mu=kT\xi=\epsilon_F\left[1-{\pi^2\over12}\left({kT\over \epsilon_f}\right)^2- {1\over2}{\lambda\over a}\left({kT\over \pi\epsilon_f}\right)^{1/2} \left(1+{d\over6}\right)\right],\eqno{(4.5)}$$
which is the desired result. From the fermion distribution function $n_\epsilon= [ 1 + e^{\beta(\epsilon-\mu)}]^{-1}$ and the above equation we see that the ${\it finite~site~effect}$  ($a <\infty$) and anisotropy effect (if $d >0$) are to 
decrease the chemical potential and thus decrease the fermion distribution function. This behavior may be interpreted as follow: The increase in the ``statistic repulsion potential" among the fermion gas by these effects, as shown in the next section, will spread particles into different states and thus decrease the fermion distribution function. 

\section{Statistical Interparticle Potential}
It is well known that, in comparison with the normal statistical behavior, bosons exhibit a larger tendency of bunching together, i.e., a positive statistical correlation. In contrast, fermions exhibit a negative statistical correlation. Uhlenbeck presented an interesting way of stating this property by introducing a ``statistical interparticle potential" and then treating the particles classically [13]. This section will present an example to show how the curvature anisotropy effects the statistical interparticle potential, as an attempt to give a ``statistical interpretation" of the results found in Sets. III and IV. 

Let us first briefly describe the method [3,13]. Define the one-particle matrix element of the Boltzmann factor by 
$$F_{ij} = <X_i|e^{-\beta H}|X_j>,\eqno{(5.1)}$$ 
then the diagonal part of the matrix element of the Boltzmann factor for a system of two identical particles can be written as 
$$ <X_1,X_2|e^{-\beta H}|X_1,X_2>= F_{11}F_{22}\pm F_{12}F_{21},\eqno{(5.2)}$$ 
where the plus (minus) sign is adopted for the boson (fermion) system. For a translation symmetry system $F_{12}$, is independent of the position and partition functions of a one-particle and two-particle system are 
$$Z_1= Tr<X_i|e^{-\beta H}|X_j> = VF_{ij}=VF_0$$
$$Z_2 =<X_1,X_2|e^{-\beta H}|X_1,X_2> =\left(VF_0\right)^2,\eqno{(5.3)}$$ 
and density matrix becomes 
$$<X_1,X_2|\tilde \rho|X_1,X_2>= {1\over Z_2}(F_{11}F_{22}\pm F_{12}F_{21})= {1\over V^2}\left[ 1\pm {F_{12}^2\over F_{0}^2}\right]. \eqno{(5.4)}$$ 
The statistical potential $v$, is defined to be such that the Boltzmann factor $exp(-\beta v)$ is precisely equal to the correlation factor (bracket term) in the above equation, i.e., 
$$v_(r) = -kT\ell n \left[1\pm {F_{12}^2\over F_{0}^2}\right]. \eqno{(5.5)}$$ 
The problem reduces to show that the factor $F_{12}^2\over F_{0}^2$ will be increased by the finite size factor and curvature anisotropy (if $d > 0$). 

Using the mode solution derived by Hu [10] we have 
$$ F_0= \sum_{n=1}^\infty\sum_{q=0}^{n-1} n~ exp\{-\beta'[n^2+d(n-1-2q)^2]\}\hspace{3cm}$$
$$\approx\sum_{n=1}^\infty\sum_{q=0}^{n-1} n~ exp(-\beta'n^2) -nd(n-1-2q)^2\beta' exp(-\beta'n^2)$$
$$=\left[1+{d\over3}\beta'\left(1+{\partial\over\partial\beta'}\right)\right]\sum_{n=1}^\infty n^2~ exp(-\beta'n^2).\hspace{2cm}. \eqno{(5.6)}$$
The last summation in . (4.1) is executed in Appendix B and we have 
$$F_0 = \left[1+{d\over3}\beta'\left(1+{\partial\over\partial\beta'}\right)\right]\left[{\pi^{1/4}\over4}(\beta')^{-3/2}\left(1-{8\pi^2\over\beta'}e^{-\pi^2/\beta'}\right)\right]$$
$$= {\pi^{1/4}\over4}(\beta')^{-3/2}\left[1-{8\pi^2\over\beta'}e^{-\pi^2/\beta'}+d \left({-1\over2}+{1\over3}\beta'\right)\right].\eqno{(5.7)}$$
To evaluate $F_{12}$ we consider the matrix element with the states $\psi_1=\psi_2=\theta_1=\theta_2=\phi_1=0$ while $\phi_2=\pi$. (The other matrix element is difficult to calculate.) Using the mode solution derived by Hu [10] we have 
$$F_{12} = \sum_{n=1}^\infty\sum_{q=0}^{n-1} n~ exp\{-\beta'[n^2+d(n-1-2q)^2]\}exp\left[{i\over2}(n-1-2q)\theta\right]$$
$$\approx \sum_{n=1}^\infty\sum_{K=-J}^{J} n~ exp(-4d\beta'K^2) exp(-\beta'n^2)exp(iK\theta)$$
$$= \sum_{n=1}^\infty n e^{-\beta'n^2}[1-d\beta'(n^2-1)]cos\left[{2\pi\over2}(n-1)\right]$$
$$=\left[1+d\beta'\left(1+{\partial\over\partial\beta'}\right)\right]\sum_{n=1}^\infty n e^{-\beta'n^2}cos\left[{2\pi\over2}(n-1)\right].\eqno{(5.8)}$$
The last summation in Eq. (4.1) is executed in Appendix B and we have 
$$F_{12} = \left[1+d\beta'\left(1+{\partial\over\partial\beta'}\right)\right]\left[\left({\pi\over\beta'}\right)^{3/2} {1\over8}e^{-\pi^2/16\beta'}(1-3e^{\pi^2/2\beta'})\right]$$
$$={1\over8}e^{-\pi^2/16\beta'}\left[1-3e^{\pi^2/2\beta'}+d\left({-1\over2}+\beta'\right) \right].\eqno{(5.9)}$$
Thus 
$${F_{12}\over F_0}\approx e^{-\pi^2/16\beta'}\left(1+{8\pi^2\over\beta'}e^{\pi^2/2\beta'}+{2\over3}d\beta'\right),\eqno{(5.10)}$$
which explicitly shows that the factor ${F_{12}\over F_0}$ will be increased by the finite size factor and curvature anisotropy (if $d > 0$). This means that in the oblate (prolate) configuration the effect of curvature anisotropy is to enhance (reduce) both the ``statistic attraction" among the bosons and the ``statistical repulsion" among fermions. 

We hope that further thermodynamic quantities, in addition to those derived in Sets. III and IV, can be interpreted by the statistical interparticle potential. 
%%%%%%%%%%%%%%%%%%%%%%%%
\section{Discussions}
In the present article we study the phenomena of Bose-Einstein condensation in a static Taub universe and find that in the oblate configuration (i.e., $d > 0$) the effect of curvature anisotropy is to increase the fraction of condensation. We also present a possible way to see that the effect of curvature anisotropy (if $d >0$) is to decrease the fermion distribution function at low temperature. We have seen that effects of curvature anisotropy on the thermodynamic quantities of the ideal gas for the universe in the prolate configuration (i.e., $d < 0$) are contrary to that in the oblate configuration (i.e., $d > 0$). As an attempt to give a ``statistic interpretation" of those found thermodynamic behaviors we have also evaluated the density matrix of a two-particle system and from this we defined the ``statistical interparticle potential" along the idea of Uhlenbeck [13]. We see that in the oblate (prolate) configuration the effect of curvature anisotropy is to enhance (reduce) both the ``statistic attraction" among the bosons and ``statistical repulsion "among the fermions. Although our investigation is confined to the nonrelativistic system, we expect that these properties will also be shown in the relativistic system. To confirm this property, a detailed analysis of the thermodynamics of a relativistic gas in the static Tau space-time is necessary. This is more difficult in mathematics and we hope to be able to report it elsewhere. 
%%%%%%%%%%%%%%%%%%%%%
~
\\
~
\\
~
\\
{\bf APPENDIX A: Calculations of the summation in Eq.(4.1)}

Using the Poisson summation formula 
$${1\over2}f(0) +\sum_{n=1}^\infty f(n) = \int_0^\infty f(t)dt + 2 \sum_{m=1}^\infty\int_0^\infty f(t) \cos(2m\pi t) dt,\eqno{(A.1)}$$
the sum in Eq. (4.1) can be written as 
$$Z_0=\sum_{n=1}^\infty {n^2\over1+e^{\beta'n^2-\xi}} = \int_0^\infty {t^2\over1+e^{\beta't^2-\xi}}dt+ 2 \sum_{m=1}^\infty {t^2 cos(2m\pi t)\over1+e^{\beta't^2-\xi}}dt\hspace{4cm}$$
$$ = \left({1\over\beta'}\right)^{3/2} \left[ \int_0^\infty {x^{1/2}\over2(1+e^{x-\xi}}dx+ \sum_{m=1}^\infty {x^{1/2}\over1+e^{x-\xi}} cos\left[2m\pi\left({x\over\beta'}\right)^{1/2}dx\right]\right]$$
$$ \equiv\left({1\over\beta'}\right)^{3/2}\left[L_0+ \sum_{m=1}^\infty L_m\right]\eqno{(A.2)}$$
For the case of $\xi>>1$, which is that at a low temperature, the integration in the above equation can be evaluated with the help of the generalized Sommerfeld's lemma [18] 
$$\int_0^\infty {g(x)\over1+e^{x-\xi}}dx=\int_0^\xi g(x) dx +{\pi^2\over6}\left({dg\over dx}\right)_{x=\xi}+{7\pi^4\over360}\left({d^3g\over dx^3}\right)_{x=\xi}+....\eqno{(A.3)}$$
Here g(x) is any well-behaved function. Thus 
$$L_0 = {1\over2}\xi^{3/2}\left(1+{\pi^2\over8}{1\over\xi^2}\right).\eqno{(A.4)}$$
$$L_m \approx \sum_{m=1}^\infty x^{1/2} cos\left[2m\pi\left({x\over\beta'}\right)^{1/2}\right]dx\hspace{11cm}$$
$$ = \sum_{m=1}^\infty {(\beta')^{3/2}\over m^3 \pi^3}\left[m\pi\left({\xi\over \beta}\right)^{1/2}\cos\left[2m\pi \left({\xi\over \beta}\right)^{1/2}\right]+\pi^2 m^2 \left({\xi\over \beta}\right) -{1\over2}\sin\left[2m\pi \left({\xi\over \beta}\right)^{1/2}\right] \right]$$
$$\equiv {1\over2}(\beta')^{1/2}\xi\,\hspace{11cm}\eqno{(A.5)}$$
and we obtain the desired result 
$$Z_0= {1\over3}\left({\xi\over\beta'}\right)^{3/2}\left[1+{\pi^2\over8}{1\over\xi^2}+{3\over2}\left(1+{d\over6}\right)\left({\beta'\over\xi}\right)^{1/2} \right].\eqno{(A.6)}$$
\\
~
\\
{\bf APPENDIX B: Calculations of the summation in Eqs.(5.6) and (5.8)}

Using the Poisson summation formula (Al) we have 
$$S=\sum_{n=1}^\infty exp(-\beta'~n^2)$$
$$= \int_0^\infty exp(-\beta'~t^2)dt + 2 \sum_{m=1}^\infty\int_0^\infty exp(-\beta'~t^2) \cos(2m\pi t) dt$$
$$ ={1\over2}\left({\pi\over\beta'}\right)^{1/2}\left(1+2\sum_{m=1}^\infty e^{-m^2\pi^2/\beta'}\right),\hspace{4cm}$$
$$\approx {1\over2}\left({\pi\over\beta'}\right)^{1/2} (1+2 e^{-\pi^2/\beta'}),\hspace{5.5cm}\eqno{(B.1)}$$
which is a suitable approximation if $a>>\lambda$ [Note that  $\beta' =(1/4\pi)(\lambda/a)$.] The summation in Eq.(5.6) is calculated from 
$$\sum_{n=1}^\infty n^2 exp(-\beta'~n^2) = -{\partial S\over\partial\beta'}.\eqno{(B.2)}$$
Using the Poisson summation formula (Al), the sum in Eq. (5.8) can be written as 
$$\sum_{n=1}^\infty n^2 exp(-\beta'~n^2)\cos\left[{\pi\over2}(n-1)\right]$$
$$= \int_0^\infty t~e^{-\beta'~t^2}\sin\left({\pi\over2}t\right)dt + 2 \sum_{m=1}^\infty\int_0^\infty t~e^{-\beta't^2}\sin\left({\pi\over2}t\right) \cos(2m\pi t) dt$$
$$ = {1\over8} \left({\pi\over\beta'}\right)^{3/2}\left[e^{-\beta'~\pi^2/16\beta'}+ \sum_{m=1}^\infty (1-4m)exp\left[-\left({\pi\over4}-m\pi\right)^2/\beta'\right]\right. $$
$$\left.+  (1+4m)exp\left[\left({\pi\over4}+m\pi\right)^2/\beta'\right]\right]\hspace{4cm}$$
$$={1\over8} \left({\pi\over\beta'}\right)^{3/2} e^{-\beta'~\pi^2/16\beta'} (1-3e^{-\pi^2/2\beta'}),\hspace{5cm} \eqno{(B.3)}$$
which is a suitable approximation if $a>>\lambda$. 

%%%%%%%%%%%%%%%%%%%%%%%
~
\\
~
\\
~
\\
{\bf  \Large References}
\begin{enumerate}
\item L. D. Landau and E. M. Lifshitz, Statistical Mechanics (Pergamon, London, 1968). 
\item R. P. Feynmam, Statistical Mechanics (Addison-Wesley, London, 1972). 
\item R. K. Pathria, Statistical Mechanics (Pergamon, London, 1972). 
\item M. B. Al'taie, J. Phys. A 11, 1603 (1978). 
\item C. A. Aragio, de Carvalho, and S. Goulart Rosa, Jr., J. Phys. A 13,989 (9180); S. Singh and R. K. Pathria, ibid. 17.2983 (1984). 
\item L. Parker and Y. Zhang, Phys. Rev. D 44, 2421 (1991). 
\item D. J. Toms, Phys. Rev. D 47, 2483 (1993). 
\item G. Cognola and L. Vanzo, Phys. Rev. D 47,4547 (1993). 
\item K. Shiraishi, Prog. Theor. Phys. 77, 975 (1987). 
\item B. L. Hu, Phys. Rev. D 8, 1048 (1973); 9, 3263 (1974). 
\item R. Critchley and J. Dowker, J. Phys. A 14, 1943 (1981); 15, 157 (1982); T. C. Shen, B. L. Hu, and D. J. O'Connor, Phys. Rev. D 31, 2401 (1985). 
\item A. Stylianopoulos, Phys. Rev. D 40, 3319 (1989). 
\item G. E. Uhlenbeck and L. Gropper, Phys. Rev. 41, 79 (1932). 
\item CL A. AragIo de Carvalho and S. Goulart Rosa, Jr., J. Phys. A 13, 3233 (1980). 
\item H. E. Haber and H. A. Weldom, Phys. Rev. Lett. 46, 1497 (1981); Phys. Rev. D 25, 502 (1982). 
\item J Bernstein and S. Dodelso, Phys. Rev. Lt. 66, 683 (1991); K. Benson, J. Bernstein, and S. Dodelso, Phys. Rev. D 44, 2480 (1992). 
\item E. C. Titchmarsh, Introduction to the Theory of Fourier Integral (Oxford University, London, 1948). 
\item A. Sommerfeld, 2. Phys. 47, 1 (1928); see also the Appendix E in Ref. 3. 
\end{enumerate}
\end{document}